# Multi-Path Routing and Wavelength Assignment (RWA) Algorithm for WDM Based Optical Networks

P. Sakthivel[1], P. Krishna Sankar[2]

[1]*Final year M.E (CSE),* [2]*Assistant Professor*
*Department of Computer Science and Engineering*
*K S R Institute for Engineering and Technology*
*Tiruchengode, Tamilnadu, India*

*Abstract-* In optical WDM networks, transmission of information along optical lines is advantageous since it has high transmission capacity, scalability, feasibility and also high reliability. But large amount of information is being carried; any problem during transmission can lead to severe damage to the data being carried. Hence it is essential to consider the routing as well as the wavelength assignment problems and then develop a combined solution for both the problems. In this paper, we propose to develop a routing and wavelength assignment algorithm for selecting the suitable alternate path for the data packets transmission. Two stages are based on the available bandwidth and the number of wavelength used in the link as construction of alternate paths, route and wavelength selection. In proposed work, Adaptive Routing and First-Fit Wavelength Assignment (AR-FFWA) algorithm to be used. For each pair of source and destination, the path with the minimum granularity values are selected as the primary path for data transmission, allocating the sufficient wavelength and the performances will be evaluated by using ns-2 simulation models. When we compared to existing system the overall blocking probability will be reduced to too low.

## I. INTRODUCTION

The basic property of single mode optical fiber is its enormous low-loss bandwidth of several tens of Terahertz. However, due to dispersive effects and limitations in optical device technology, single channel transmission is limited to only a small fraction of the fiber capacity [1]. To take full advantage of the potential of fiber, the use of wavelength division multiplexing (WDM) technology has become the option of choice. With WDM, a number of distinct wavelengths are used to implement separate channels. An optical fiber can carry several channels in parallel, each on a particular wavelength. The number of wavelengths that each fiber can carry simultaneously is limited by the physical characteristics of the fiber and the state of the optical technology used to combine these wavelengths onto the fiber and isolate them off the fiber. With currently available commercial technology, a few tens of wavelengths can be supported within the low-loss. The main characteristics of WDM can concisely be summarized as follows:

- Fully photonic network where fiber amplifiers are used
- Several channels are transmitted simultaneously in each fiber
- The network forms a wide backbone-network

Optical networks employing wavelength division multiplexing (WDM) offer the promise of meeting the high bandwidth requirements of emerging communication applications, by dividing the huge transmission bandwidth of an optical fiber (~50 terabits per second) into multiple communication channels with bandwidths (~10 gigabits per second) compatible with the electronic processing speeds of the end users.

*A. Wavelength Division Multiplexing (WDM)*

Optical wavelength division multiplexing (WDM) networking is being recognized as an efficient technique for upcoming wide area network environments, because of its impending ability to meet the increasing demands of low latency communication and high bandwidth [2]. WDM allows several signals to be carried independently along the same fiber provided each signal uses a different wavelength. As a result same fiber can be shared by many connections.

*B. Advantages of Optical Networks*

- Optical fiber Networks have high capacity
- Restoration done in optical layer is faster and efficient
- Optical signals can be transferred as long distance
- Wide Bandwidth
- Light Weight, Small Diameter

There are three different types of wavelength division multiplexing:

- WDM (Wavelength Division Multiplexing)
- CWDM (Coarse Wavelength Division Multiplexing)
- DWDM (Dense Wavelength Division Multiplexing)

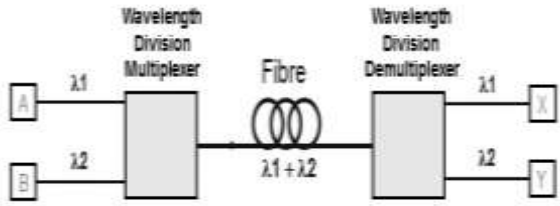

Fig. 1. Wavelength division multiplexing

Fig. 1 will shows how the wavelength will be assigned to the each path of the channels and each wavelength is like a separate channel (fiber).

*C. Routing and Wavelength Assignment (RWA)*

The routing and wavelength assignment problem is an optical networking problem with the goal of maximizing the number of optical connections [3]. Each connection request must be given a route and wavelength. The wavelength must be consistent for the entire path, unless the usage of wavelength converters is assumed. Two connections requests can share the same optical link, provided a different wavelength is used [4].

*D. Routing*

*1) Fixed Routing:* In fixed routing, a single fixed route is predetermined for each source-destination pair. When a connection request arrives, the network will attempt to establish a light-path along the fixed route. If no common wavelength is available on every link in the route, then the connection will be blocked. A fixed routing approach is simple to implement; however, it is very limited in terms of routing options and may lead to a high level of blocking. In order to minimize the blocking in fixed routing networks, the predetermined routes need to be selected in a manner which balances the load evenly across the network links. Fixed routing schemes do not require the maintenance of global network state information.

*2) Fixed-Alternate Routing:* In fixed-alternate routing, each node in the network is required to maintain a routing table which contains an ordered list of number of fixed routes to each destination node. When a connection request arrives, the source node attempts to establish the connection on each of the routes from the routing table in sequence, until a route with a valid wavelength assignment is found. If no available route is found from the list of alternate routes, then connection request is blocked and lost. In most cases, the routing tables at each node are ordered by the number of fiber link segments (hops) to the destination. Therefore, the shortest path to the destination is the first route in the routing table. When there are ties in the distance between different routes, one route may be selected at random. Another advantage of fixed-alternate routing is that it can significantly reduce the connection blocking probability compared to fixed routing.

*3) Adaptive routing:* In adaptive routing, the route from a source node to a destination node is chosen dynamically, depending on the network state. The network state is determined by the set of all connections that are currently in progress. One form of adaptive routing is adaptive shortest cost path routing. For the network in Figure 2, if the links (1,2) and (4,2) in the network are busy, then the adaptive routing algorithm can still establish a connection between Nodes 0 and 2, while both the fixed routing protocol and the fixed-alternate routing protocols with fixed and alternate paths would block the connection.

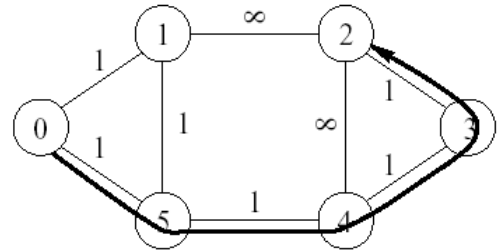

Fig. 2. Adaptive routing

Fig. 2 will shows how the adaptive routing will selects the suitable path.

*E. Wavelength Assignment Algorithm*

In routing there are two ways to use the wavelength. In First case, we can use the same wavelength throughout the path which spans from source to destination [5]. In the second case, we are allowed to use different wavelengths in the path running from source to destination. Among possible others, few independent wavelength assignment algorithms are,

- Random (R) wavelength assignment
- First-Fit (FF) assignment
- Most-used (MU) assignment
- Least-Used (LU)

## II. EXISTING SYSTEM

The O-OFDM networks are only using the Single path routing so the overall bandwidth blocking probability will high [9]. If congestion occurs in between routing path that data could be discard from the transmission process [10]. Here more data loss, more blocking probability and low through put. An O-OFDM network are using Single Path Routing and Breadth-First Path Search (BFPS) algorithm used. The Single-path routing can causes unbalanced traffic distributions, When Traffic load is high, difficulty to find a single path routing algorithm [11] [12].

*A. Worst Case of Existing System*

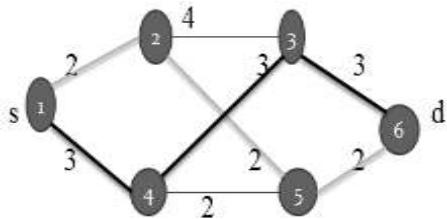

Fig. 3. Fixed Path Routing in Existing System

Fig. 3 will shows fixed path routing for the source node 's' to destination node 'd' and here the Bandwidth (BW) allocation as BW=2 and BW=3 for entire transmission only the fixed route will be used.

### III. PROPOSED SYSTEM

The dynamic routing processes of optical networks are implemented for these proposed systems [13]. It will be implemented by using RWA algorithm for selecting route and wavelength assignments and RMSA with HSMR algorithm to be used for reducing the bandwidth blocking probability [14]. The previous Optical system to be considering the static path transmission but now it will propose and develop this dynamic path routing and wavelength assignment algorithm for WDM Optical network. It is increase the Throughput, reduces the blocking probability and time delay. The Adaptive routing path approach to be used in proposed system. Alternate path routing assumes that every node stores the first N shortest and predetermined paths to each destination by utilizing global and local node link information. Congestion is determined in part by the number of wavelengths available per path or link. It is fully dynamic one; those are realized in simulation results.

This proposed work, a route between a sender node and receiver node is selected from a set of Pre-defined alternate routes. If any collision occurs in the routed path that time the data will be path changed and transmit alternate path from source to destination [15]. Multi-Path routing provides increased throughput and utilizes the network resources more efficiently. In the proposed work, a route between a sender node and a receiver node is selected from a set of pre-defined routes which are link-disjoint paths, i.e., paths which do not share a link, and a wavelength is selected along the selected route [16]. Then the proposed work selects a route among pre-defined paths and assigns wavelengths to segments along the selected route in such a way as to avoid the generation of bottleneck links and the depletion of a specific wavelength. In this case, further light-paths cannot be established in the link. Therefore, it expects to reduce the blocking probability by avoiding the generation of bottleneck links. In order to distribute loads and avoid the generation of bottleneck links, the proposed scheme preferentially selects a route which has segments with many available wavelengths [17] [18].

### IV. SYSTEM FLOW DIAGRAMS

Fig. 4 will shows the system flows of proposed works.

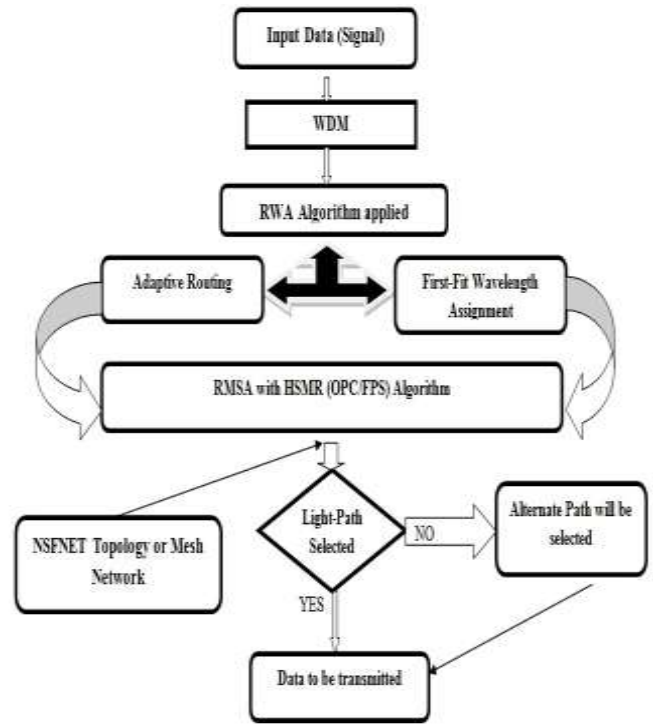

Fig. 4. System flow diagrams of Optical networks

Figure 4 has shown the modulated optical signals with data to be transmitted through optical fiber, here the RWA algorithm an Adaptive routing and First-Fit wavelength to be assigned to each path of the networks. Finally, light-path should be established by using RMSA with HSMR (OPC/FPS) algorithm and then the data to be transmitted in different routes [19]. If the congestion will occurs in light path immediately alternate path should be assigned for the congested path. Then the performances are evaluated using NSFNET topology for the simulations [20].

### V. SIMULATION RESULTS

In this paper, we are using Network Simulator (ns2) for simulating the proposed Adaptive Routing and First-Fit Wavelength Assignment (AR-FFWA) algorithm for performance evaluation [6] [7] [8]. This will reduce the overall blocking probability and increasing the throughput.

*A. Simulation Modules*

1. Data packets will be sent to the destination node by the source node.
2. The possibility of occurrence of congestion in the transmission route.
3. The Alternate Path (backup path) will be choosing and the traffic will be routed through backup path to avoid collision.
4. Performance Analysis

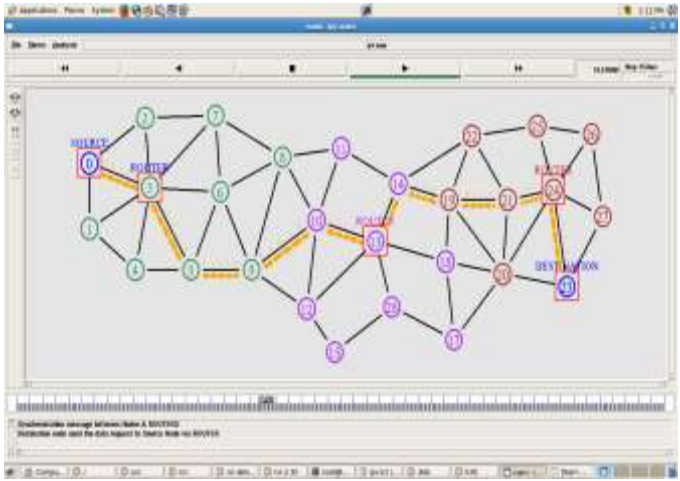

Fig. 6. Destination node can send the request to the Source node via the Routers

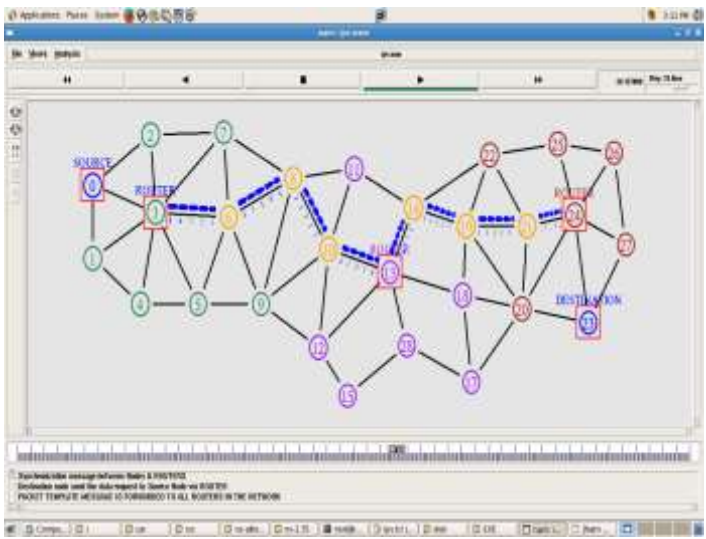

Fig. 7. Packet Template messages to be send to the destination node

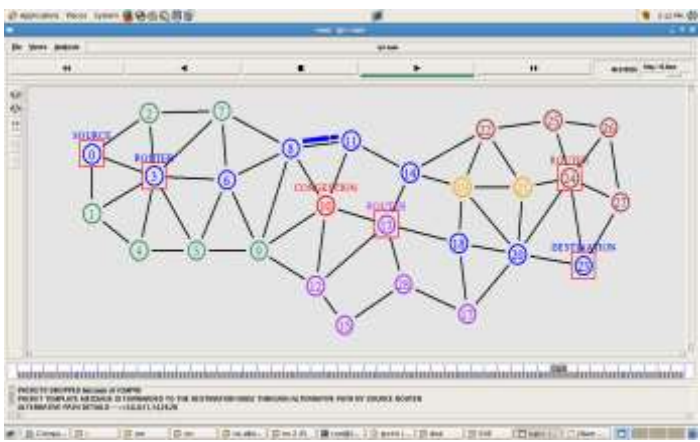

Fig. 8. Congested node can be identified in the transmission route and data can be transmitted via the alternate paths

Fig. 6, 7 and 8 can shows the data transmission, congestion identification with the alternate path selection.

TABLE I
PARAMETERS USED IN DIFFERENT NETWORK TOPOLOGY DESIGN

| Traffic Loads (Erlangs) | Total costs of the Path |
|---|---|
| *8 node Network topology* | |
| *Granularity =1* | |
| 300 | 9 |
| 400 | 12 |
| 500 | 11 |
| *Granularity =2* | |
| 100 | 11 |
| 200 | 12 |
| *12 node Network topology* | |
| *Granularity =3* | |
| 700 | 26 |
| 800 | 18 |
| 900 | 31 |
| *14 node Network topology* | |
| *Granularity =3* | |
| 500 | 9 |
| 600 | 32 |
| 700 | 37 |

Table 1. Will explains that different traffic loads (Erlangs) with total cost of the different paths and below the graph models can explain simulation results.

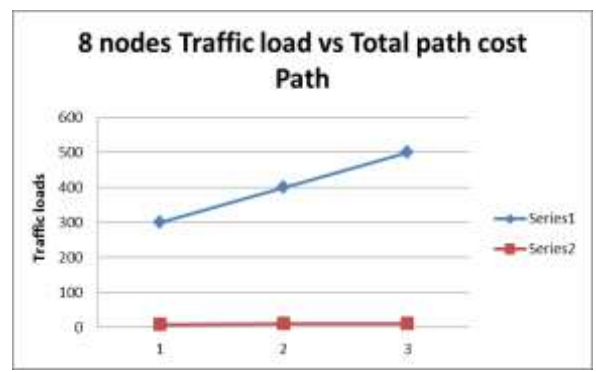

Fig. 9. Distribution of the path-distance requests to the different paths and traffic loads with granularity=1 in 8 node network topology

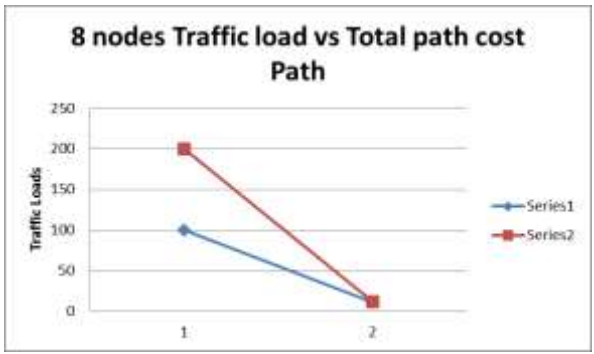

Fig. 10. Distribution of the path-distance requests to the different paths and traffic loads with granularity=2 in 8 node network topology

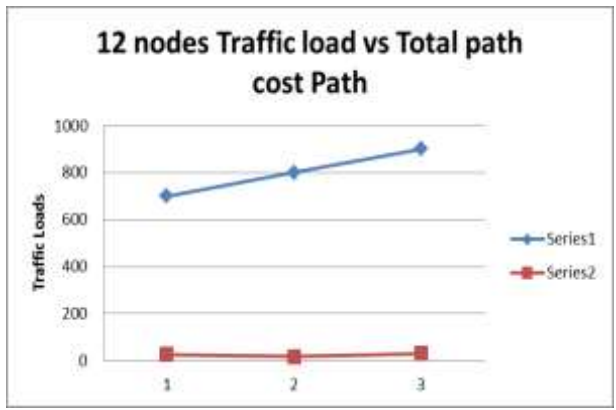

Fig. 11. Distribution of the path-distance requests to the different paths and traffic loads with granularity=3 in 12 node network topology

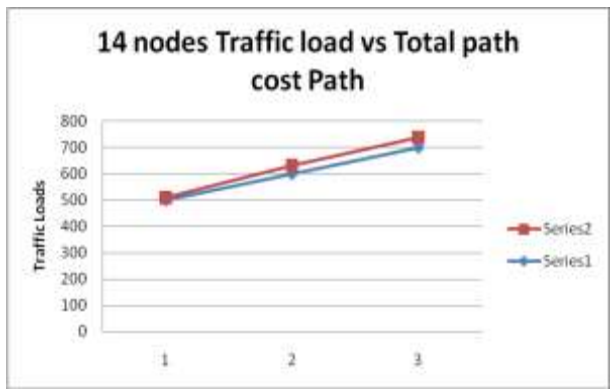

Fig. 12. Distribution of the path-distance requests to the different paths and traffic loads with granularity=3 in 14 node network topology

Fig. 9,10,11 and 12 illustrates the distribution of the path-distance requests between the different paths and traffic loads and number of paths for provisioned connection requests using MP-ARFF with granularity value as g=1, 2, 3. When the traffic loads when the traffic load is lower than 900 Erlangs, all requests are served with a multiple routing path. From 400 to 1000 Erlangs, the percentage of multi-path provisioned connections increases gradually [22] [23]. Here 8 nodes, 12 node and 14 nodes network topology can be simulated with MP-ARFF.

## VI. CONCLUSIONS

This paper proposed a dynamic RWA scheme for WDM optical networks with routing and wavelength assignment algorithm. The route and wavelengths are selected for each light-path based on wavelength availability, Adaptive Routing and First-Fit Wavelength Assignment (AR-FFWA) algorithm to be used. In order to establish an efficient path for a given source and destination pair, initially the bandwidth and delay will be calculated for every link which can form a path between the given node pair. Incase if the link is completely loaded then, it will not be further considered, incase if it not completely loaded and has free channels for wavelength assignment then it is further considered for the formation of the path. Next the shortest and low traffic path is calculated based on assigning minimum granularity values. Furthermore, it will demonstrated the robustness of the projected scheme against system parameter values such as the number of wavelengths supported by each fiber and the number of fibers in each link. Finally this proposed work developed the dynamical RWA algorithm for optical networks for fast and long distance data transmission. Through simulation experiments, observed that the proposed scheme efficiently reduces blocking probability in WDM optical networks.